\documentclass{ifacconf}

\makeatletter
\let\old@ssect\@ssect %
\makeatother

\usepackage{graphicx}      
\usepackage{natbib}       

\usepackage{amsmath}
\usepackage{dsfont}
\usepackage{hyperref} 

\makeatletter
\def\@ssect#1#2#3#4#5#6{%
  \NR@gettitle{#6}%
  \old@ssect{#1}{#2}{#3}{#4}{#5}{#6}%
}
\makeatother

\begin{document}
\begin{frontmatter}

\title{Non-linear State-space Model Identification from Video Data using Deep Encoders}

\author[First]{Gerben I. Beintema} 
\author[First,Second]{Roland Toth} 
\author[First]{Maarten Schoukens}

\address[First]{Department of Electrical Engineering, Eindhoven University of Technology, 5600 MB, Eindhoven, The Netherlands (e-mails: g.i.beitema@tue.nl, r.toth@tue.nl, m.schoukens@tue.nl).}
\address[Second]{Systems and Control Laboratory, Institute for Computer Science and Control, Kende u. 13-17, H-1111 Budapest, Hungary.}

\begin{abstract}                %
Identifying systems with high-dimensional inputs and outputs, such as systems measured by video streams, is a challenging problem with numerous applications in robotics, autonomous vehicles and medical imaging. In this paper, we propose a novel non-linear state-space identification method starting from high-dimensional input and output data. Multiple computational and conceptual advances are combined to handle the high-dimensional nature of the data. An encoder function, represented by a neural network, is introduced to learn a reconstructability map to estimate the model states from past inputs and outputs. This encoder function is jointly learned with the dynamics. Furthermore, multiple computational improvements, such as an improved reformulation of multiple shooting and batch optimization, are proposed to keep the computational time under control when dealing with high-dimensional and large datasets. We apply the proposed method to a video stream of a simulated environment of a controllable ball in a unit box. The study shows low simulation error with excellent long term prediction capability of the model obtained using the proposed method.
\end{abstract}

\begin{keyword}
Non-linear State-Space Modelling, Deep Learning, Pixels, Multiple Shooting.
\end{keyword}

\end{frontmatter}

\section{Introduction}

Systems with high dimensional inputs and outputs (i.e. large-scale systems) are ever more prevalent due to the increased presence of, for instance, high-resolution video cameras, PDE simulations, system networks, and medical imaging devices. Hence, the identification of flexible models and methods for modelling nonlinear large-scale systems is of the essence. However, currently, this is a challenging task due to the curse of dimensionality and the difficulty of modelling nonlinearities that are encountered in these systems~\citep{moerland2020survey-RL-model-based}.

There is extensive literature available for linear state-space model identification for large-scale systems such as subspace methods~\citep{Overschee2012subspace-lin-book}, expectation-maximization~\citep{Gibson2005maximum-linelihood}, and PCA or CCA~\citep{Katayama2007subspacebook}. However, non-linear state-space identification for large-scale systems is currently an open problem. 

Recent results for non-linear state-space identification present considerable advances in, state estimation~\citep{courts2020variational}, polynomial state-space models \citep{decuyper2020unconstrained-multiple-shooting}, and artificial neural networks based state-space models~\citep{schoukens2020NL-LFR,masti2018autoencoder,mavkov2020integrated}. Furthermore, parameter estimation methods for non-linear state-space models have improved considerably by the introduction of the multiple shooting method with considerable theoretical~\citep{ribeiro2019multiple-shooting} and practical results~\citep{decuyper2020unconstrained-multiple-shooting}. These models and estimation methods have yet to be analysed and developed for large-scale systems. 

One successful approach to identify non-linear large-scale systems combines a non-linear autoencoder for dimension reduction with an multiple input multiple output (MIMO) NARX model~\citep{wahlstrom2015pixel-IO-autoencoder} (this approach will be referred to as ``IO autoencoder'' within this paper). The IO autoencoder approach outperforms linear identification methods and allows for model predictive control~\citep{wahlstrom2015pixel-IO-autoencoder-MPC}. However, a MIMO NARX model is considerably more difficult to interpret and to use for controller design than non-linear state-space models. The complexity of a MIMO NARX model also rapidly increases for growing dynamical complexity. Furthermore, the NARX model structure often degrades in performance when used for simulation. 

The aim of this paper is to develop an encoder-informed non-linear state-space identification approach that can efficiently process high-dimensional input-output data. To this end this paper combines \textit{i)} non-linear state-space models parameterized as artificial neural networks, \textit{ii)} a non-linear encoder together with \textit{iii)} an improved formulation of the multiple shooting method utilizing batch optimization. Here the non-linear encoder enables the identification of large-scale systems. The proposed method only requires a single loss function, obtains state-of-the-art results using randomly initialized model parameters and allows for simulation error minimization.\footnote{Implementation of the proposed method is available at \url{https://github.com/GerbenBeintema/SS-encoder-video}}

The paper is structured as follows: Section 2 provides an overview of the proposed method, Section 3 shows the application of the proposed method to a numerical example followed by the conclusions in Section 4.

\section{The state-space encoder method}

\subsection{Model structure}
We aim to estimate the following discrete-time state-space model: 
\begin{subequations}
\begin{align}
    \hat{x}_{t+1} &= f_\theta(\hat{x}_t,u_t), \\
    \hat{y}_t &= h_\theta(\hat{x}_t),
\end{align}
\end{subequations}
with $t \in \mathds{Z}$ the time index, $\hat{y}_t \in \mathds{R}^{n_X \times n_Y}$ the model output, $y_t$ the system output, $\hat{x}_t \in \mathds{R}^{n_x}$ the internal model state, $u_t \in \mathds{R}^{n_u}$ the input, $\theta$ the model parameters and $f_\theta$ and $h_\theta$ the state and output function. Artificial Neural Networks (ANN) are used to represent $f_\theta$ and $h_\theta$ as they have excellent approximation properties for high-dimensional functions \citep{Barron1993}. We assume that the measured data is generated by a system contained within this model class: $y_t = h_{\theta_0}(x_t,u_t) + v_t$ and $x_{t+1} =  f_{\theta_0}(x_t,u_t)$, where a possibly coloured, additive zero-mean finite-variance noise source $v_t \in \mathds{R}^{n_y}$ is assumed to be present at the system output.

\begin{figure}[t]
    \centering
    \includegraphics[width=1\linewidth]{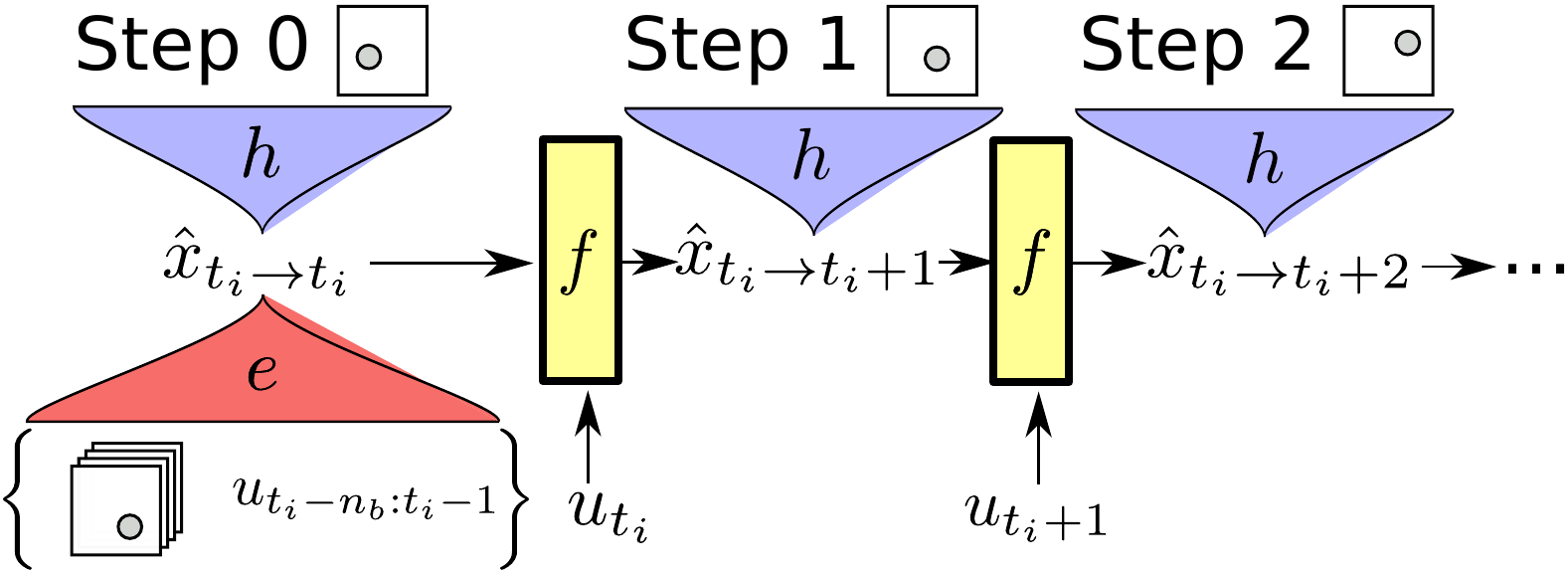}
    \caption{The proposed non-linear state-space model estimation method where the initial state $\hat{x}_{t_i \xrightarrow{} t_i}$ is estimated by a state encoder function $e$ based on previous measured input samples and output frames.}
    \label{fig:encoder}
\end{figure}

\subsection{Parameter estimation}
Most commonly, non-linear state-space models with an OE noise structure are estimated by minimizing the simulation loss (i.e. $V_{\text{sim}}(\theta) \sim \sum_t ||h_\theta(x_t) - y_t||^2_2$), however, the computational cost scales linearly with the number of samples $O(N_{\text{samples}})$. 

To improve the scalability of the proposed method with the length of the dataset, the proposed loss function is constructed by summing over $N$ sub-sections with starting indices $t_i$ and length $T+k_0+1$, similar to the multiple shooting method~\citep{ribeiro2019multiple-shooting}, as: 
\begin{subequations}
\begin{align}
    V_{\text{encoder}}(\theta) &= \frac{1}{2 N (T+1)} \sum_{i=1}^N \sum_{k=k_0}^{T+k_0} ||\hat{y}_{t_i \xrightarrow{} t_i + k} - y_{t_i+k}||^2_2, \\
     \hat{y}_{t_i \xrightarrow{} t_i + k} &:= h_\theta( \hat{x}_{t_i \xrightarrow{} t_i +k}),\\
     \hat{x}_{t_i \xrightarrow{} t_i + k+1} &:= f_\theta(\hat{x}_{t_i \xrightarrow{} t_i + k},u_{t_i + k}),\\
     \hat{x}_{t_i \xrightarrow{} t_i} &:= e_\theta(y_{t_i-n_a:t_i-1},u_{t_i-n_b:t_i-1}),
     \label{eq:encoder}
\end{align}
\end{subequations}
where $x_{t_i \xrightarrow{}  t_i + k}$ indicates $k$ recursive uses of $f_\theta$ to calculate the state as: 
\begin{gather*}
    \hat{x}_{t_i \xrightarrow{} t_i + k} = f_\theta(f_\theta( ... f_\theta(\hat{x}_{t_i \xrightarrow{} t_i}, u_{t_i}),..., u_{t_i + k-2}), u_{t_i + k-1}).
\end{gather*}

The initial state $\hat{x}_{t_i \xrightarrow{} t_i}$ is given by an encoder function $e_\theta$ based on the previous input and output samples $u_{t-n_b:t-1} \in \mathds{R}^{n_u \cdot n_b}$ and $y_{t-n_a:t-1} \in \mathds{R}^{(n_X \times n_Y) \cdot n_a}$ which in fact estimates a reconstructability map of the underlying nonlinear system hence, we call the proposed method the state-space encoder method. It is graphically presented in Figure \ref{fig:encoder}. Just like the state and output functions $f_\theta$ and $h_\theta$, the encoder function $e_\theta$ is represented as an ANN to ensure excellent approximation properties when dealing with high-dimensional data \citep{Barron1993}.

The proposed method resolves some of the shortcomings of the parametric start method~\citep{decuyper2020unconstrained-multiple-shooting} (i.e. $\hat{x}_{t_i \xrightarrow{} t_i}$ is introduced as a parameter of the model) for it has a fixed model complexity whereas the parametric start scales linearly with the number of sections. Moreover, the encoder can act as an observer even on unseen data which, for instance, can jump-start simulations with the approximately correct internal state and aid in control.

\begin{figure}
    \centering
    \includegraphics[width=0.5\linewidth]{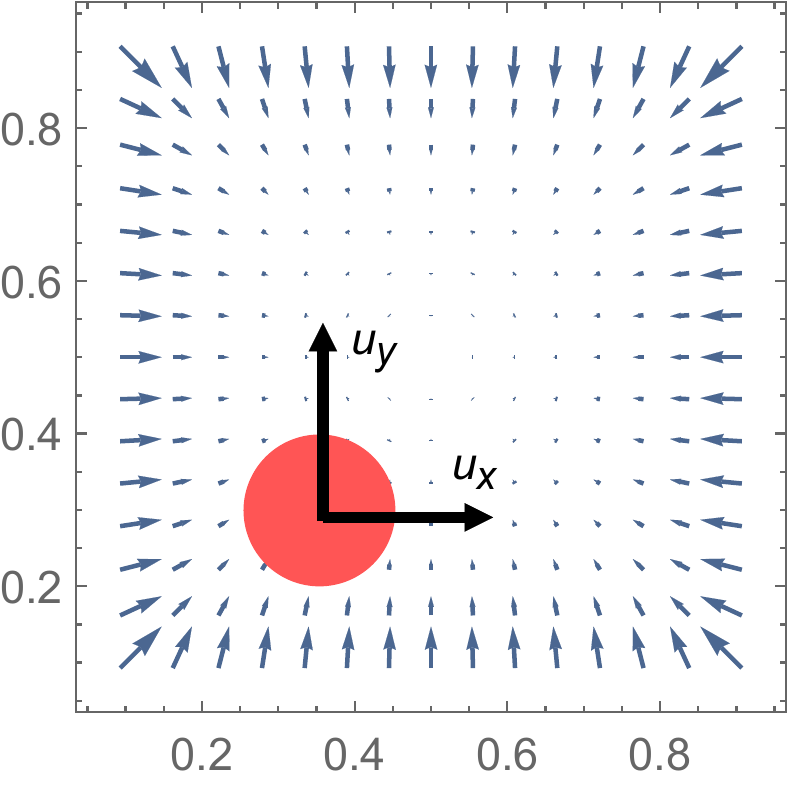}
    \caption{The considered numerical environment that consists of a ball contained within a square unit box with strong non-linear repulsive forces near the four boundaries and background friction. The actuation (input $u$ of the system) applies forces on the ball in both directions ($u_x$, $u_y$) and the video output consists of a 25 by 25 pixels array per frame.  }
    \label{fig:ball-scheme}
\end{figure}

\begin{figure*}[ht]
    \centering
    \includegraphics[width=1\textwidth]{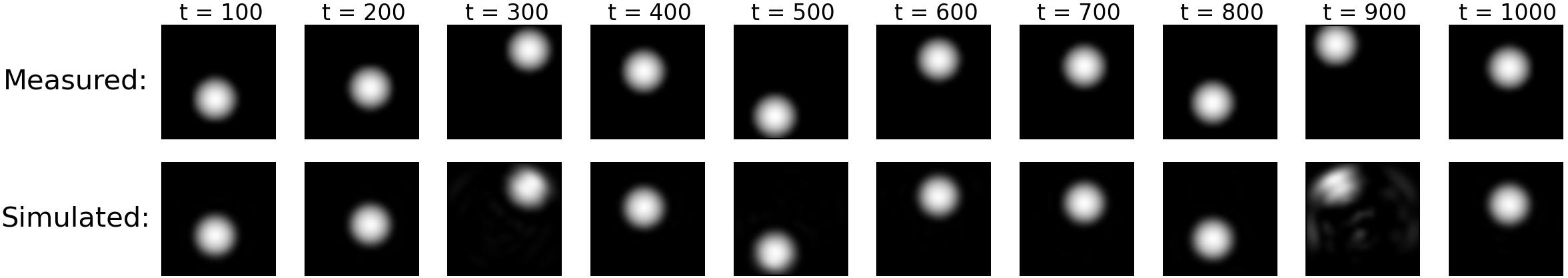}
    \caption{Output frames of both the measured output of the systems and the simulated outputs of the state-space encoder method for different instances of time in the test set. (Supplementary Video: \url{https://youtu.be/L_zCwzpMZUc}) }
    \label{fig:strip}
\end{figure*}

\begin{figure*}[ht]
    \centering
    \includegraphics[width=1\textwidth]{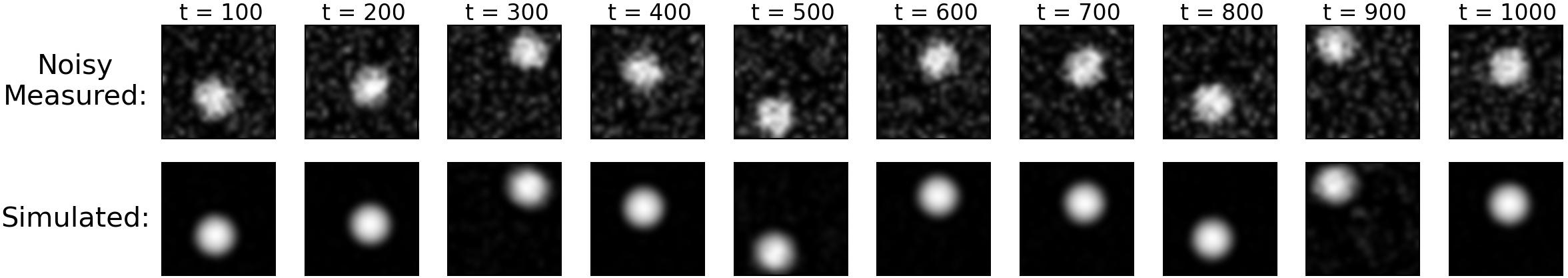}
    \caption{Similar to Figure \ref{fig:strip}, but with added noise to the camera output with a noise to signal ratio of $a=100\%$ (i.e. 0 dB) (see Eq. \ref{eq:noise-added}) . It can be seen that the identified non-linear state-space model is able to correctly identify the system and reduce the noise to a negligible level. (Supplementary Video: \url{https://youtu.be/kPXVluTZ-i0})}
    \label{fig:strip-noise}
\end{figure*}

Furthermore, due to the independence of the loss function on each section, the proposed method allows for \textit{i)} computational speedup by utilizing modern parallelization methods and \textit{ii)} the utilization of batch optimization methods. The batch formulation of the state-space encoder method is obtained by summing not over all sections, but only a subset $\mathcal{B}$ of section as 
\begin{subequations}
\begin{align}
    V_{\text{batch}}(\theta) &= \frac{1}{2 N_{\text{batch}} (T+1)} \sum_{i \in \mathcal{B} } \sum_{k=k_0}^{T+k_0} ||\hat{y}_{t_i \xrightarrow{} t_i + k} - y_{t_i+k}||^2_2, \\
    \mathcal{B} &\subset \{1,2,...,N\}.
\end{align}
\end{subequations}
This reformulation can utilize modern powerful batch optimization algorithms developed by the machine learning community (e.g. Adam~\citep{kingma2014adam}). Furthermore, utilizing the batch formulation only requires the data to be partially loaded which can be necessary for large data sets of large-scale systems where memory constraints play an essential role. 

\section{Numerical Experiments}
\subsection{Simulated Environment}

As an example of a large-scale system, we consider a simulated environment with a camera of 25 by 25 pixels pointed at a ball which can move within a unit box graphically shown in Figure \ref{fig:ball-scheme}. This environment can be interpreted physically as a ball in a concave square cup which can be tilted at different angles to force the ball to move and a camera observing the situation from above. More precisely we consider the following non-linear ODE for the dynamics of the ball
\begin{subequations}
\begin{gather}
    \ddot p_x = \beta \left (\frac{1}{p_x^2}- \frac{1}{(1-p_x)^2} \right ) - \gamma \dot p_x + k u_x   \\
    \ddot p_y = \beta \left (\frac{1}{p_y^2}- \frac{1}{(1-p_y)^2} \right ) - \gamma \dot p_y + k u_y
\end{gather}
\end{subequations}
where $(p_x,p_y)$ is the position of the ball, $(u_x,u_y)$ the external forces bounded by $-1\leq u_x \leq 1$ and $-1\leq u_y \leq 1$, $\beta = 1/200$, $\gamma = 0.79$, $k=1/4$, an initial position of $p_x(0) = p_y(0) = 0.5$ with zero initial velocity. We discretize time by integrating the system with the time step $\Delta t = 0.3$ using the Runge–Kutta method and the zero-order hold condition for the actuation.  

To calculate the pixel intensity at a position $X,Y$ we use the following non-linear equation
\begin{equation}
    \text{pixel}(X,Y) = \max \left ( 0, 1-\frac{(X-p_x)^2+(Y-p_y)^2}{r^2} \right ) + v.
\end{equation}
where $r=0.25$ and $v$ is a noise term. The video image $y_t$ is obtained by sampling this continuous image in a uniform $n_X = 25$ by $n_Y = 25$ grid ranging from 0 to 1 in both directions.

We perform three separate experiments where the inputs are sampled from a uniform distribution $u_x,u_y \sim \mathcal{U}(-1,1)$ compatible with the input constraints. The three experiments are i) a noisy training set of size 30,000, ii) a noisy validation set of size 5000 and iii) a noiseless test set of size 5000 frames. Multiple instances of the same training and validation data are created with different amplitudes of added Gaussian noise as 
\begin{equation}
\label{eq:noise-added}
    v \sim \mathcal{N}(0, (\sigma_y a)^2)
\end{equation} 
over each image where $a$ is the noise to signal ratio and $\sigma_y = 0.204$ the standard deviation of the test set.

\subsection{State-space encoder approach}

We implement the state-space encoder as follows; we utilize the same feedforward neural network structure for $e_\theta$, $f_\theta$ and $h_\theta$ consisting of a two hidden layer neural network, 64 nodes per layer, $\tanh$ activation and a linear bypass. The parameters are initialized by sampling the uniform distribution $\mathcal{U}(-\sqrt{k},\sqrt{k})$ with $k=1/\sqrt{n_{\text{in}}}$ where $n_{\text{in}}$ is the number of inputs to the layer. The images are flattened into $25 \cdot 25 = 625$ vectors for both the input of the encoder and the output of $h_\theta$.

We choose the following state-space encoder hyperparameters of $n_x = 6$, $k_0 = 0$, $T = 50$, $n_a = n_b = 5$. The state dimension $n_x$ is chosen deliberately a bit larger than the minimal state dimension of 4 for we noticed that the optimization is significantly faster (by approximately 25\%) and more consistent. We suspect the mechanism behind this is similar to immersion which reduces the complexity of the required functions by increasing the state dimension~\citep{ohtsuka2005immersion}. Furthermore, the multiple shooting starting indices $t_i$ are allowed to be every possible value within the range of the training set. For training, the Adam batch optimization method~\citep{kingma2014adam} is used with the default learning rate of $\alpha = 10^{-3}$ and a batch size of 256. After each epoch, the simulation error is calculated on the validation set and the current model is saved if a new lowest simulation loss has been achieved. Furthermore, we normalize both the inputs and outputs by subtracting the mean and dividing by the standard deviation, this improves performance and training time. 

The obtained results of the method proposed in this paper are compared with a reproduction of the IO autoencoder~\citep{wahlstrom2015pixel-IO-autoencoder} approach which is implemented utilizing the same neural network structure and as similar as possible hyperparameters (i.e. $n_z = n_x = 6$, $n = n_a = n_b = 5$). 

\subsection{Results}

We applied the state-space encoder method on the ball-in-box simulated camera system as mentioned in the previous section. The performance of the state space encoder can be compared visually by observing the output frames of a simulation run on the test set as seen in Figure \ref{fig:strip} for the zero noise case. The difference is minimal for most instances of time which suggest that our proposed method can accurately find a non-linear state-space model of order 6 for this system. Moreover, For the case with additive measurement noise, the proposed method is still able to successfully model the system as shown in Figure \ref{fig:strip-noise}. 

Quantitatively the proposed method has an NRMS simulation error lower than the IO autoencoder method as can be seen in Table \ref{tab:test-perform} (the NRMS error is obtained by dividing the RMS error by the standard deviation of the output in th test set $\sigma_y = 0.204$). In the same table, it can be seen that even relatively high values of noise $a$ only marginally decrease the performance of the proposed method whereas the IO autoencoder degrades significantly faster. We suspect that this difference is due to the OE noise structure present in the considered system. 

\begin{table}[]
\centering
\caption{NRMS of the simulation error on the noiseless test set for various levels of noise in the training data.}
\begin{tabular}{c|cc}
\multicolumn{1}{l|}{Noise Level $a$} & \multicolumn{1}{l}{\begin{tabular}[c]{@{}l@{}}IO autoencoder\\ \citep{wahlstrom2015pixel-IO-autoencoder}\end{tabular}} & \multicolumn{1}{l}{\begin{tabular}[c]{@{}l@{}} State-space Encoder\\ (Proposed Method)\end{tabular}} \\ \hline
0\%                              & 7.35\%                                                                                                                                  & 6.57\%                                                                                             \\
5\%                              & 7.61\%                                                                                                                                  & 7.07\%                                                                                             \\
20\%                             & 8.56\%                                                                                                                                  & 7.55\%                                                                                             \\
50\%                             & 12.20\%                                                                                                                                 & 7.90\%                                                                                             \\
100\%                            & 14.04\%                                                                                                                                 & 9.40\%                                                                                            
\end{tabular}
\label{tab:test-perform}
\end{table}

\begin{figure}[!b]
    \centering
    \includegraphics[width=1\linewidth]{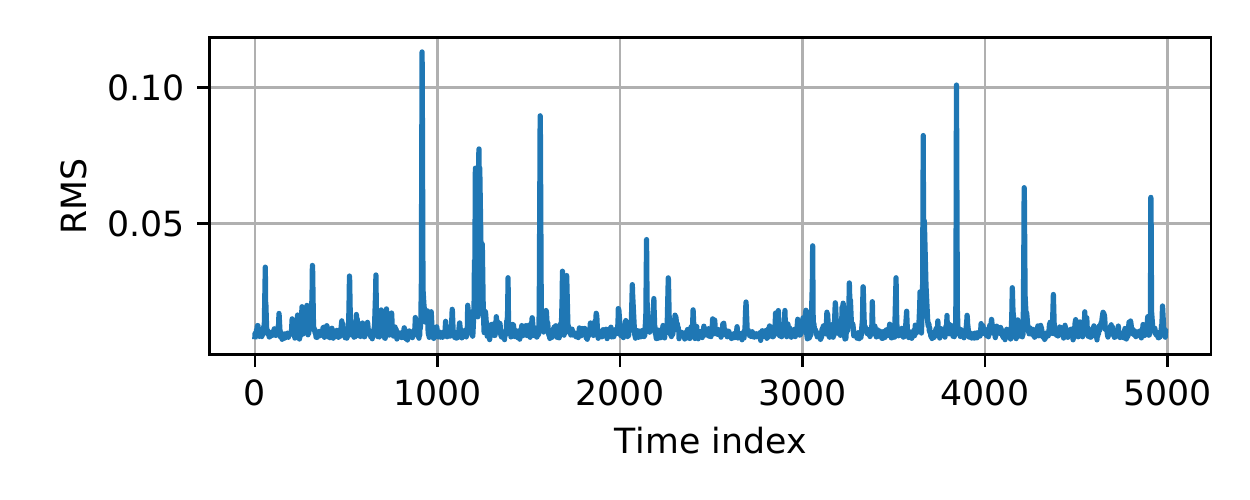}
    \caption{The root mean square error of the simulated output of each output frame (i.e. $||y_t - \hat{y}_t||_2/\sqrt{n_X n_Y}$) for the state-space encoder applied on the test set. The large spikes are instances where the model is used in regions of the state-space were no training data is present.}
    \label{fig:time-error}
\end{figure}

However, looking closely at the simulations on the test set shows instances of time where the error becomes relatively high. This can also be observed in the root mean square error of a single output frame over time as in Figure \ref{fig:time-error} where there are spikes of high simulation error go up to ten times the average. After close inspection, we found that these spikes appeared when the model was used in regions where there was little to no training data available and hence requiring extrapolation. This can be resolved by having larger training sets or using input design methods which aim to fill the state-space more efficiently \citep{Heinz2017}. 

\begin{figure}
    \centering
    \includegraphics[width=1\linewidth]{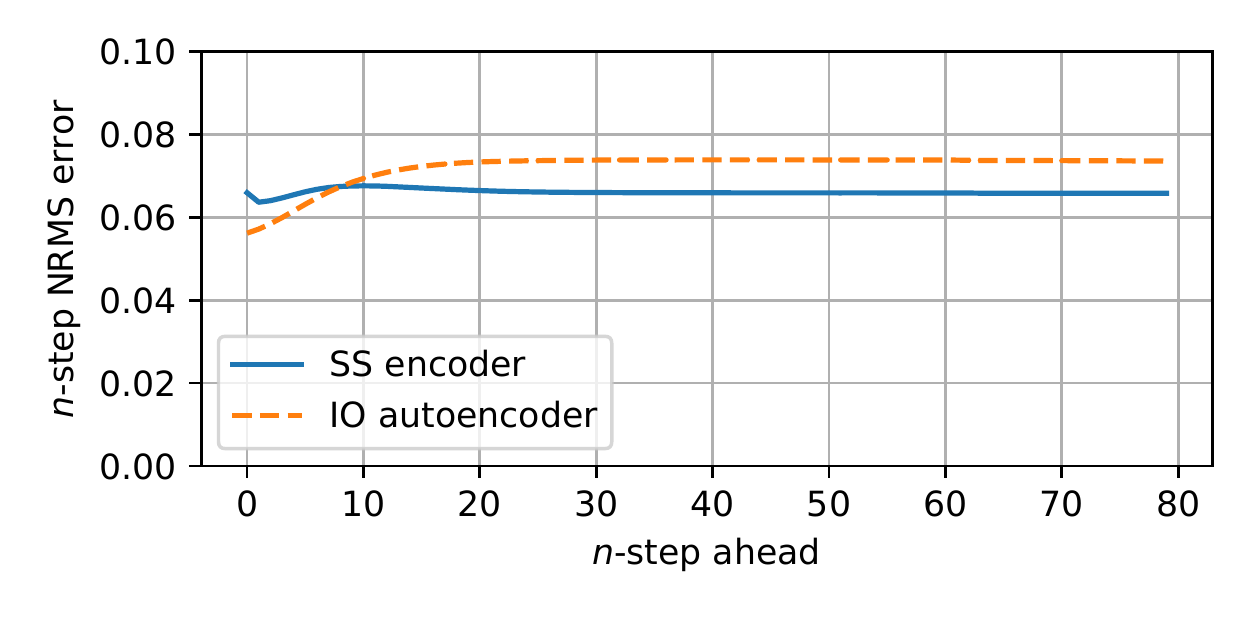}
    \caption{The n-step NRMS error of the model output response (Eq. \eqref{eq:n-step-NRMS}) on the test set which shows improved long term prediction of the state-space encoder compared to the IO auto-encoder.}
    \label{fig:n-step-error}
\end{figure}

Lastly, to measure the effectiveness of the encoder and validate the chosen hyperparameters of the encoder method we use the $n$-step NRMS. This is the expected normalized $n$-steps ahead prediction error after the encoder presumed state estimate. More formally:
\begin{align}
\label{eq:n-step-NRMS}
    \text{NRMS}_n = \frac{\sqrt{1/M \sum_{i=1}^{M} (\hat{y}_{t_i \xrightarrow{} t_i + n} - y_{t_i + n})^2} }{\sigma_y}.
\end{align}
This quantity is shown in Figure \ref{fig:n-step-error} where the average is taken over all the possible starting indices $t_i$ of the test set. For the state-space encoder method, the $n$-step NRMS is almost a straight line which indicates that the encoder is able to estimate the initial state accurately with these hyperparameters, whereas, the IO autoencoder degrades slightly in performance for increasing $n$ which is typical for the NARX model structure as it minimizes the one step ahead prediction error.

\section{Conclusions}
This paper proposed a novel method to effectively identify non-linear state-space models for large-scale systems. The proposed method utilizes a state-space encoder, a reformulation of the multiple shooting method which allows for the batch optimization to be used. This results in a optimization approach that scales well for increasing input-output dimensions and data set length. The method was able to identify with high accuracy a ball system with non-linear dynamics and video stream output while being less affected by increased levels of noise compared to the IO autoencoder. This was achieved without explicitly modelling the dynamics of the ball and purely learning from pixels and force inputs. Applications of this method to physical systems and a detailed theoretical analysis are subject of future work.

\bibliography{references}
\end{document}